\documentclass[prl,twocolumn,showpacs,amsmath,amssymb]{revtex4}
\usepackage{graphicx}
\usepackage{dcolumn}
\usepackage{bm}
\newcommand{\be}{\begin{equation}}
\newcommand{\ee}{\end{equation}}
\newcommand{\bea}{\begin{eqnarray}}
\newcommand{\eea}{\end{eqnarray}}
\newcommand{\eps}{\varepsilon}
\newcommand{\ev}[1]{\langle#1\rangle}
\newcommand{\ddt}{\frac{\partial}{\partial t}}

\newcommand{\ihddt}{i\hbar\frac{\partial}{\partial t}}
\newcommand{\mcal}[1]{{\mathcal{#1}}}
\newcommand{\mrm}[1]{{\mathrm{#1}}}
\newcommand{\drm}{{\mathrm{d}}}
\newcommand{\e}{\mathrm{e}}

\renewcommand{\vec}[1]{{\mathbf{#1}}}

\newcommand{\pv}{\vec{\epsilon}}
\begin{document}
%
\title{Photoluminescence and Terahertz Emission from Femtosecond\\
Laser-Induced Plasma Channels}
%
\author{W.~Hoyer}
\email{hoyer@acms.arizona.edu}
\author{J.V.~Moloney}
\author{E.M.~Wright}
\affiliation{%
Arizona Center for Mathematical Sciences and Optical Sciences
Center, University of Arizona, Tucson, AZ 85721, USA}
\author{A. Knorr}
\affiliation{%
Institut f\"ur Theoretische Physik, AG Nichtlineare Optik und
Quantenelektronik, Technische Universit\"at Berlin, 10623 Berlin,
Germany}
\author{M.~Kira}
\author{S.\,W.~Koch}
\affiliation{%
Department of Physics and Material Sciences Center,
Philipps-University, Renthof 5, D-35032 Marburg, Germany}
%
\date{\today}
%
\begin{abstract}
Luminescence as a mechanism for terahertz emission from
femtosecond laser-induced plasmas is studied. By using a fully
microscopic theory, Coulomb scattering between
electrons and ions is shown to lead to luminescence even for a
spatially homogeneous plasma. The spectral features
introduced by the rod geometry of laser-induced plasma channels
in air are discussed on the basis of a generalized mode-function
analysis.
\end{abstract}

\maketitle
%
%
%
The past decade has seen increasing activity in the generation and
detection of far-infrared or terahertz (THz) radiation with
accompanying increase in applications, including time-domain THz
spectroscopy, and THz tomography \cite{Davies:04}. Current THz sources
have limited spatial extent ($\mu$m to mm), such as transient
antenna emitters, sources based on optical rectification, and
quantum cascade lasers.  A notable exception is the observation of
THz radiation emitted from plasma channels induced in air by
femtosecond (fs) infrared filaments
\cite{Proulx:00,Tzortzakis:02,Mechain:03}, or light strings, that
can extend over centimeters
\cite{Nibbering:96,Tzortzakis:03,Yu:03,Ladouceur:01}. The
extension of these results to other media offers the possibility
of a new variety of extended THz sources.

In addition to the technological importance of light strings for
realizing THz sources, the mechanism of THz emission raises basic
physics issues.  To date most theory and experiments on {\em
coherent} THz emission in air have considered localized plasmas
with volumes $\simeq$ $\mu$m$^3$ \cite{Hamster:93}, much less than
$\lambda_{\mrm{THz}}^3$ for typical THz wavelengths. In contrast, light
string induced plasmas can be significantly longer than
$\lambda_{\mrm{THz}}$ along the propagation direction of the exciting
laser meaning that coherent emissions from different
locations in the plasma need not add constructively \cite{Sprangle:04}.
In this Letter, we propose {\it incoherent} light emission by a
two-component plasma as a viable mechanism for THz emission since
this does not rely on the establishment of relative phases between
emissions from different locations in the plasma. We develop a
microscopic theory which includes the Coulomb interaction between
the charge carriers and the light-matter interaction with a
quantized light field. First we show that even a homogeneous
plasma can produce broad-band radiation in the form of
luminescence due to the electron-ion scattering. Second we
make a mode function analysis for the specific rod-like geometry
for light string induced plasmas in air in order to explore the
angular dependence of the THz emission.

%
Our starting point is the total Hamiltonian
%
\be H_{\mrm{tot}} =  \sum_{\lambda } \left(H_{\mrm{kin}}^{\lambda}
                             +  H_{A\cdot p}^{\lambda}
                             +  H_{A^2}^{\lambda}
                  \right)
                             + H_{\mrm{L}}+ H_{\mrm{C}} ,
\label{eq:Htot} \ee
%
where $\lambda=e,i$ sums over electron and ion contributions, and
we next explain the meaning of each of these terms. This form of
the total Hamiltonian arises from the minimal substitution
Hamiltonian $H^{\lambda}_{\mrm{min}} = \int \drm^3r
\Psi^{\dagger}_{\mrm{\lambda}}(\vec{r}) \frac{\left(
\hat{\vec{p}} - q_\lambda \vec{A}(\vec{r})
\right)^2}{2m_\lambda} \Psi_{\mrm{\lambda}}(\vec{r})$ applied to
our system \cite{Cohen:89}, where
$\Psi_{\mrm{\lambda}}(\vec{r})$ and
$\Psi^{\dagger}_{\mrm{\lambda}}(\vec{r})$ are the annihilation
and creation operators for electrons and ions,
respectively, $m_\lambda$ and $q_\lambda$ are the particle masses
and charges, $ \vec{A}(\vec{r}) = \sum_{\vec{q},\sigma}
(\mcal{A}_{q} B_{\vec{q},\sigma}\e^{i\,\vec{q}\cdot\vec{r}}
\pv_{\vec{q},\sigma} + \mbox{h.c.})$ is the quantized vector
potential operator. Here $\mcal{A}_{q}=\sqrt{\hbar/(2\eps_0
\omega_q\,V)}$, and we introduced the boson operators
$B_{\vec{q},\sigma} (B^{\dagger}_{\vec{q},\sigma})$ which
annihilate (create) photons with wave number $\vec{q}$ and
polarization direction $\pv_{\vec{q},\sigma}$, in terms of which the free-field
Hamiltonian appearing in Eq.~(\ref{eq:Htot}) is $H_{\mrm{L}} =
\sum_{\vec{q},\sigma} \hbar\omega_{q} \left(
B^{\dagger}_{\vec{q},\sigma} B_{\vec{q},\sigma} +\frac{1}{2}
\right)$.  We work in the $\vec{A}\cdot \vec{p}$-picture and
formulate our equations in momentum space, the relation between
real and momentum space operators being $\Psi_{\lambda}(\vec{r})
= 1/\sqrt{V} \sum_{\vec{k}} \e^{i\,\vec{k}\cdot\vec{r}}
a_{\lambda,\vec{k}},$ where the operator $a_{\lambda,\vec{k}}$ annihilates a
particle of species $\lambda$ with momentum $\hbar \vec{k}$.
The kinetic energy term in the Hamiltonian (\ref{eq:Htot}) is then
given by $H_{\mrm{kin}}^{\lambda} = \sum_{\vec{k}} \eps^{\lambda}_{k}\,
a^{\dagger}_{\lambda,\vec{k}} a_{\lambda,\vec{k}}$, with free-particle
energies $\eps^{\lambda}_{k}=\frac{\hbar^2 k^2}{2\,m_\lambda}$,
and the light-matter interaction terms become
%
\be
H_{A\cdot p}^{\lambda}
=
-\sum_{\vec{k},\vec{q}}
\vec{J}^{\lambda}_{\vec{k}}\cdot \vec{A}_{\vec{q}} \,
         a^{\dagger}_{\lambda,\vec{k}+\vec{q}/2}
         a_{\lambda,\vec{k}-\vec{q}/2},
\label{eq:HAp_e}
\ee
%
with the canonical current matrix element
$\vec{J}^{\lambda}_{\vec{k}} = \frac{q_\lambda}{m_\lambda}\hbar\vec{k},$
and the Fourier transformed operator of the vector potential $\vec{A}_{\vec{q}}$,
along with the nonlinear $A^2$-contribution
%
\be
H_{A^2}^{\lambda}
=
\frac{q_\lambda^2}{2\,m_\lambda} \sum_{\vec{k},\vec{q},\vec{q}',s}
\vec{A}^{\dagger}_{\vec{q}'} \cdot \vec{A}_{\vec{q}}\,
a^{\dagger}_{\lambda,\vec{k}+\vec{q},s} a_{\lambda,\vec{k}+\vec{q}',s}.
\label{eq:HA2_e}
\ee
%
Finally, Coulomb interactions between charges are incorporated
in the total Hamiltonian~(\ref{eq:Htot}) using
%
\be H_{\mrm{C}} = \frac{1}{2} \sum_{\lambda,\lambda' \atop
\vec{k},\vec{k}',\vec{q}} V_{q}^{\lambda,\lambda'} \,
a^{\dagger}_{\lambda,\vec{k}} a^{\dagger}_{\lambda',\vec{k}'}
a_{\lambda',\vec{k}'+\vec{q}} a_{\lambda,\vec{k}-\vec{q}} .
\label{eq:HC} \ee
%
with the Coulomb matrix element $V_{q}^{\lambda,\lambda'}$ \cite{Haug:94}.

We work in the Heisenberg picture where the quantum average of
an operator $O$ evolves according to $i\hbar\,\partial\ev{O}/\partial
t=\ev{[O, H_{\mrm{tot}}]}$. Here, the many-body
Coulomb and the light-matter interaction introduce a hierarchy
problem such that e.g.\ photon numbers $\ev{B^{\dagger} B}$ 
are coupled to mixed correlations of
$\ev{B^{\dagger} a^{\dagger} a}$ which in turn become coupled to
higher-order expectation values like $\ev{B^{\dagger} a^{\dagger}
a^{\dagger} a a}$. We truncate this hierarchy using a cluster
expansion \cite{Wyld:63,Hoyer:04b} where multi-particle or mixed
expectation values are consistently factorized. For example,
photon correlation functions are split according to
%
\be
\ev{B^{\dagger}_{\vec{q}',\sigma'} B_{\vec{q},\sigma}}
= \ev{B^{\dagger}_{\vec{q}',\sigma'}} \ev{B_{\vec{q},\sigma}}
 +\Delta\ev{B^{\dagger}_{\vec{q}',\sigma'} B_{\vec{q},\sigma}} ,
\ee
%
with $\ev{B_{\vec{q},\sigma}}$ related to the coherent component
of the field and $\Delta\ev{B^{\dagger}_{\vec{q}',\sigma'}
B_{\vec{q},\sigma}}$ describing quantum fluctuations.

Here we are interested in incoherent light emission under
quasi-stationary conditions after the exciting laser has passed,
and an appropriate measure of the photoluminescence is the total
rate of energy emitted into a single mode $(\vec{q},\sigma)$ given
by
%
\be I_{\vec{q},\sigma}^{\mrm{PL}}=\hbar \omega_{q}
\ddt\ev{B^{\dagger}_{\vec{q},\sigma} B_{\vec{q},\sigma}} = \hbar
\omega_{q} \ddt\Delta\ev{B^{\dagger}_{\vec{q},\sigma}
B_{\vec{q},\sigma}} , \ee
%
where we have neglected any coherent field contribution. The
Heisenberg equation of motion for the photon number is found to be
%
\bea
\lefteqn{
\ddt \Delta \ev{B^{\dagger}_{\vec{q},\sigma}
B_{\vec{q},\sigma}}
=}
\nonumber\\
&&\quad
- \frac{2}{\hbar} \mrm{Im}
\left[
\sum_{\vec{k}} \vec{J}^{e}_{\vec{k}} \cdot
\pv^*_{\vec{q},\sigma} \, \mcal{A}_{q}
\Delta\ev{B^{\dagger}_{\vec{q},\sigma}
e^{\dagger}_{\vec{k}-\vec{q}/2} e_{\vec{k}+\vec{q}/2}}
\right],
\label{eq:ddtB+B_diag} \eea
%
where we have defined $ e_{\vec{k}} \equiv a_{\lambda=e,\vec{k}}$ for
notational simplicity. Here we have neglected the $A^2$-contribution
in order to investigate the general emission mechanism. We will show
later that its influence can be accounted for by generalized mode functions
of the quantized field such that all our calculations remain valid.
From Eq.~(\ref{eq:ddtB+B_diag}) we see that the incoherent emission of
photons is driven by photon-assisted densities of the form
$\Delta\ev{B^{\dagger}_{\vec{q},\sigma}
e^{\dagger}_{\vec{k}-\vec{q}/2} e_{\vec{k}+\vec{q}/2}}$, which
correspond to processes where an electron changes its momentum,
while emitting a photon with the corresponding momentum difference.
Under the assumption of large ion masses we have neglected a 
similar contribution from photon-assisted ion densities 
$\Delta\ev{B^{\dagger} p^{\dagger} p}$. To evaluate the dynamics
of $\Delta\ev{B^{\dagger}_{\vec{q},\sigma}
e^{\dagger}_{\vec{k}-\vec{q}/2} e_{\vec{k}+\vec{q}/2}}$
we apply the long wavelength and low density limit; the procedure
leads to
%
\bea
\lefteqn{
\ihddt \Delta\ev{B^{\dagger}_{\vec{q},\sigma}
e^{\dagger}_{\vec{k}-\vec{q}/2} e_{\vec{k}+\vec{q}/2} } =}
\nonumber\\
&&
\bigl(
\frac{\hbar^2 (\vec{k}\cdot \vec{q}) }{m_e}  - \hbar \omega_{q}
 - i \Gamma^{D}_{\vec{k}}(q) \bigr)
\Delta\ev{B^{\dagger}_{\vec{q},\sigma} e^{\dagger}_{\vec{k}-\vec{q}/2} e_{\vec{k}+\vec{q}/2} }
\nonumber\\
&&
+ i \sum_{\vec{k}'} \Gamma^{OD}_{\vec{k},\vec{k}'}(q)\,
\Delta\ev{B^{\dagger}_{\vec{q},\sigma} e^{\dagger}_{\vec{k}'-\vec{q}/2} e_{\vec{k}'+\vec{q}/2} }
\nonumber\\
&&
+ \mcal{A}_{q} \, \pv_{\vec{q},\sigma} \cdot \vec{J}^{e}_{\vec{k}} f^e_{\vec{k}},
\label{eq:ddtB+e+e_short}
\eea
where the last term corresponds to the spontaneous emission source
term.

It is well known that a single-component, non-interacting plasma cannot
emit or absorb radiation while simultaneously conserving energy and 
momentum. However, in Coulomb interacting systems such restrictions
do not apply. In the following, we concentrate on the scattering
between electrons and ions, symbolized by $\Gamma^{D}$ and $\Gamma^{OD}$ in
Eq.~(\ref{eq:ddtB+e+e_short}). We numerically confirmed that
emission due to electron-electron scattering is negligible compared 
to the electron-ion scattering. Additional mechanisms such
as electron-neutral scattering will increase the scattering rates but
are not expected to change the qualitative picture. A microscopic
description of the scattering is obtained by including the
coupling of the photon-assisted densities to photon-assisted
correlations of the form $\Delta\ev{B^\dagger e^\dagger p^\dagger
p e}$, deriving the Heisenberg equation of motion for this
quantity, and factorizing the resulting higher-order correlations in
products of the form $f(1-f)(1-f) \Delta \ev{B^\dagger e^\dagger
e}$. By Fourier transformation this equation can be formally
solved and inserted into Eq.~(\ref{eq:ddtB+e+e_short}) in the form of
frequency dependent scattering matrices
%
\be
\Gamma^{OD}_{\vec{k},\vec{k}'}(q) =
 \frac{(-i) W^2_{|\vec{k}-\vec{k}'|} \sum_{\vec{k}''}f^i_{\vec{k}''}}
{\eps^e_{\vec{k'}} - \eps^e_{\vec{k}} - \hbar \omega_q - i \delta}
+ \bigl\{\omega_q \leftrightarrow -\omega_q \bigr\}^*  , 
\label{eq:def_GOD_ei} 
\ee
with diagonal elements $\Gamma^{D}_{\vec{k}}(q) = \sum_{\vec{k}'}
\Gamma^{OD}_{\vec{k}',\vec{k}}(q)$, where $W_{|\vec{k}-\vec{k}'|}$ 
denotes the statically screened Coulomb matrix element \cite{Haug:94}
and large ion mass and low density limit are applied. The frequency
dependence of the scattering derives from the presence of the
photon operator in the photon assisted correlations.
Inspection of the denominator in Eq.~(\ref{eq:def_GOD_ei}) shows
that an electron changes momentum due to Coulomb interaction with an ion.
The corresponding change in kinetic energy is accounted for by photon
emission while the total change of momentum of the electron is absorbed 
by the heavy ions.

%
\begin{figure}
\rotatebox{0}{%
\resizebox{0.4\textwidth}{!}{%
\includegraphics{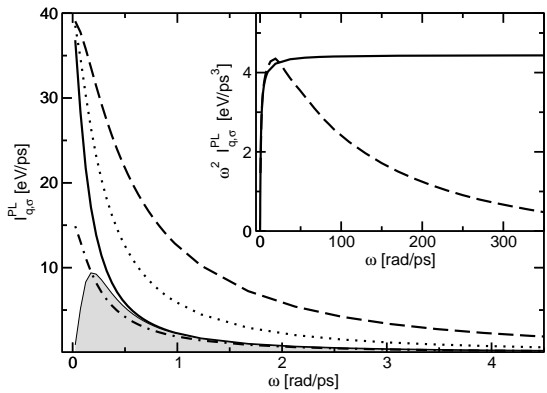}}
}
\caption{Luminescence signal $I^{\mrm{PL}}_{\vec{q},\sigma}$ as function
of $\omega_q = c q$ for a plasma of density $n=10^{16}\,\mrm{cm}^{-3}$
(solid), $n=2 \times 10^{16}\,\mrm{cm}^{-3}$ (dotted),
and $n=4 \times 10^{16}\,\mrm{cm}^{-3}$ (dashed) at $T=3000\,$K,
and for $n=10^{16}\,\mrm{cm}^{-3}$ at $T=2000\,$K (dash-dotted). The shaded
area is obtained by multiplication of the solid line with the 
mode strength $|\vec{u}_{\vec{q}}|^2$.
Inset: data for $n=10^{16}\,\mrm{cm}^{-3}$ and $T=3000\,$K after
multiplication with $\omega^2$ with (solid) and 
without (dashed) frequency dependent scattering.
}
\label{fig:PLdens}
\end{figure}
%
Figure~\ref{fig:PLdens} shows the calculated $I^{\mrm{PL}}_{\vec{q},\sigma}$
as function of frequency for a homogeneous plasma with different electron
densities and temperatures. The temperatures lie in the range of typical
excess energies of about 0.25\,eV for multi-photon ionization. We observe
that the spectra are sharply peaked around $\omega=0$. The emission at
$\omega=0$ is mainly determined by the plasma temperature while the width
of the emission tail increases with elevated density. In order to understand these
results, we approximate the microscopic ($k$-dependent) scattering matrices
by an effective constant damping rate $\gamma_{\mrm{eff}}$. In that case,
Eq.~(\ref{eq:ddtB+e+e_short}) can be formally solved and inserted
into Eq.~(\ref{eq:ddtB+B_diag}), which upon assuming
Maxwell-Boltzmann distributions for the electrons and taking the
low density limit yields
%
\bea
I_{\vec{q},\sigma}^{\mrm{PL}}
&=&
\omega_{\mrm{pl}}^2 k_{\mrm{B}}T \,
\frac{\gamma_{\mrm{eff}}}
{\omega_q^2 + \gamma_{\mrm{eff}}^2}.
\label{eq:IPL}
\eea
%
If we take the main feature of Eq.~(\ref{eq:def_GOD_ei}) and assume
that $\gamma_{\mrm{eff}}$ increases linearly with
density, we see that the high frequency luminescence is proportional
to $\omega_{\mrm{pl}}^2 \,\gamma_{\mrm{eff}} \propto n^2$ while the
low frequency behaviour is determined by 
$\omega_{\mrm{pl}}^2 /\gamma_{\mrm{eff}} \propto 1$ and thus independent
of density. For low frequencies, Fig.~\ref{fig:PLdens} also exhibits
the trend of decreasing emission strength with decreasing temperature.
The observed insensitivity of the high-frequency spectrum on temperature,
however, cannot be explained with Eq.~(\ref{eq:IPL}). Nevertheless,
this equation can be used to fit an estimate for typical scattering times.
For the case of $n=10^{16}\,\mrm{cm}^{-3}$ and $T=3000\,$K we obtain
a scattering time of the order of 2.5\,ps.

Another shortcoming of Eq.(\ref{eq:IPL}) lies in the fact that for 
high frequencies it is proportional to $1/\omega^2$ such that the
integrated power density proportional to
$\omega^2\,I^{\mrm{PL}}_{\vec{q},\sigma}$ becomes divergent. In contrast,
the microscopic, frequency dependent scattering leads to an exponential
decay of $\omega^2\,I^{\mrm{PL}}_{\vec{q},\sigma}$ as shown in the inset
of Fig.~\ref{fig:PLdens}. In order to obtain an estimate for the 
conversion efficiency of emission into the THz regime, we integrate the
total emitted power density $\omega_q^2 (2 \pi^2 c^3)^{-1} 
I_{\vec{q},\sigma}^{\mrm{PL}}$ from $\omega=0$ up to $\omega=50\,$THz
and obtain a value of roughly $\mcal{P}_{\mrm{THz}} 
\approx 5 \times 10^4$\,W/m$^3$. 
Comparing the total energy emitted by a plasma channel of radius
$R=30\,\mu$m and length of $l=1\,$m over the characteristic
lifetime $\tau=1$\,ns, $E = \pi R^2 l \tau \,\mcal{P}_{\mrm{THz}} 
\approx 1.4 \times 10^{-13}$\,J, with typical light-string excitation
energies in the range of $10$\,mJ, we obtain
a conversion efficiency of $\eta \approx 1.4 \times 10^{-11}$ comparable
to estimates for coherent emission mechanisms \cite{Sprangle:04}.

So far, we have completely neglected the influence of the $A^2$-contribution
to the Hamiltonian. Instead of including the resulting terms into the
equations of motion, we pursue a different approach. We approximate
the density operator in Eq.~(\ref{eq:HA2_e}) by its factorized expectation
and assume that the overall density profile varies slowly enough to be
taken as constant. In that case, $H_{A^2}^{e}$ is quadratic
in the field in analogy to $H_{\mrm{L}}$. The combined   $H_{A^2}^{e}+H_{\mrm{L}}$
can be diagonalized by choosing the light modes according to the 
generalized Helmholtz equation
%
\be
c^2 \,\nabla \times \nabla \times \vec{u}_{\vec{q},\sigma}(\vec{r})
+\omega_{\mrm{pl}}^2(\vec{r}) \,\vec{u}_{\vec{q},\sigma}(\vec{r})
= \,\omega^2\,\vec{u}_{\vec{q},\sigma}(\vec{r})
\label{eq:wave}
\ee
%
with the space-dependent plasma frequency $\omega_{\mrm{pl}}(\vec{r}) 
= \sqrt{\frac{e^2 n(\vec{r})}{\eps_0 m_e}}$. 
The computation of the new mode functions is a purely classical problem, the
quantum character of the light field being fully contained in the photon
annihilation and creation operators. If the extension of the plasma is
much larger compared to characteristic electronic length scales, the 
charge carriers still see an approximately three-dimensional environment.
In that case, we find only one modification that the luminescence spectrum
is multiplied by a frequency-dependent mode function.
For a thin slab of constant density with a thickness $d$ well
below typical emission wavelengths, this prefactor for emission perpendicular
to the plane of the plasma is simply given by
\[
|\vec{u}_{q}(\vec{r}=0)|^2 = \left( 1 + \frac{1}{4} \frac{\omega_{\mrm{pl}}^4 d^2}{q^2 c^4} \right)^{-1}.
\]
Thus, we see how frequencies well below the plasma frequency are effectively blocked
out. The shaded area in Fig.~\ref{fig:PLdens} shows the resulting spectrum
for an assumed thickness $d=3\,\mu$m. For thicknesses $d \gg c/\omega_{\mrm{pl}}$
all frequencies below the plasma frequencies would be completely absent in the 
emission.

The geometry of a laser induced light string enters our computations in exactly
the same way. The eigenmodes have to be computed according to 
Eq.~(\ref{eq:wave}) in the presence of the spatial density profile $n(\vec{r})$
of the plasma column\cite{THzPRL}.  Our problem has well separated length scales, namely,
the thermal wavelength of the electrons $\lambda_{\mrm{th}}\simeq
1$\,nm is much smaller than the diameter of the plasma column
$R\approx 30\,\mu$m, which is in turn much smaller than typical
THz wavelengths $\lambda_{\mrm{THz}}$. As far as optical coupling is
concerned, the exact form of $n(\vec{r})$ is therefore relatively
unimportant as long as it is strongly localized, say at the origin
of the $(x-y)$ plane. We have solved for the eigenmodes of Eq.~(\ref{eq:wave}) 
by using a transfer matrix technique, extending the approach of 
Ref.~[\onlinecite{Helaly:97}] to oblique incidence.
All solutions can be classified by the wave vector $\vec{q}$ of an
incoming plane wave and its polarization $\sigma$ distinguishing
between TM-mode (incident electric field in the plane of incidence) and
TE-mode (incident electric field perpendicular to electron string).

\begin{figure}
\resizebox{0.46\textwidth}{!}{%
\includegraphics{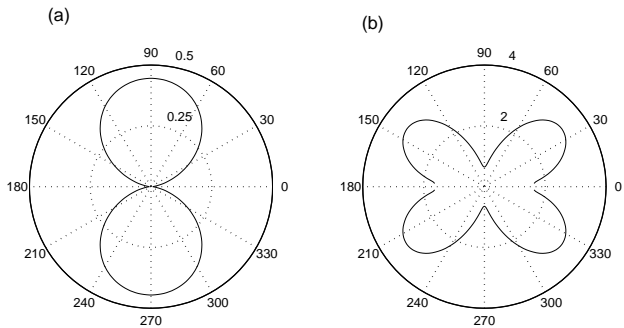}}
\caption{Mode strength $|u_{\vec{q},\sigma}(\vec{r}=0)|^2$ for
(a) $c\,q = 1\,\mrm{ps}^{-1} < \omega_{\mrm{pl}}$ and (b) $c\,q \approx 4.5\,\mrm{ps}^{-1}
= 0.8\, \omega_{\mrm{pl}}$. The corresponding mode strength in vacuum would be 1.
}
\label{fig:modestrength}
\end{figure}

The mode strength $|\vec{u}_{\vec{q},\sigma}(\vec{r}=0)|^2$ for
the TM-mode is presented in Fig.~\ref{fig:modestrength} for the
case of a step-like density profile with a carrier density of
$n=10^{16}$\,cm$^{-3}$ over a cylinder with radius $R=30\,\mu$m.
%
%
For frequencies well below the plasma frequency, we find a dipole
radiation pattern qualitatively similar to that observed
experimentally in Ref.~[\onlinecite{Tzortzakis:02}]. The main
emission for those frequencies is perpendicular to the plasma rod
aligned along the horizontal axis of Fig.~\ref{fig:modestrength}. 
This emission pattern can be traced back to the increasing 
suppression of light modes below the plasma frequency as the
emission angle deviates more from the normal direction.
For frequencies around and above the plasma frequency,  
the angle dependence drastically changes since this suppression
does not exist anymore. For
certain frequencies and emission angles resonance effects can be
observed which depend on the precise choice of parameters. 
In the given example, we observe a preferred emission angle around 40 degrees.
The mode strength of the TE-mode does not strongly depend on the angle
of incidence since for any angle the electric
field component lies perpendicular to the light string.

%
In summary, we presented a microscopic theory for the
photoluminescence emitted from fs laser-induced plasmas, and have
shown that even a homogeneous two-component plasma can emit due to
electron-ion scattering. The angle dependence of the THz emission for light
string induced plasmas was explored using modal analysis, a dipole
pattern for emission frequencies below the plasma frequency being
found in qualitative agreement with the recent experiment
\cite{Tzortzakis:02}. In contrast, a drastically different angle
dependence for emission frequencies above the plasma frequency is
predicted. Finally, we note that although our analysis was motivated
by studies of THz emission in air
\cite{Proulx:00,Tzortzakis:02,Mechain:03}, our results apply more
generally, for example, to glasses or electron-hole plasmas in
semiconductors.
%
%
\begin{acknowledgments}
We thank Marten Richter for valuable discussions.
The work is sponsored in Tucson by the U.S. Air Force Office of
Scientific Research (AFOSR), under grant AFOSR-F49620-00-1-0312,
in Berlin by the Deutsche Forschungsgemeinschaft through the SFB 296,
and in Marburg by the Optodynamics Center and the Deutsche 
Forschungsgemeinschaft.
\end{acknowledgments}
%
%

%
%

\begin{thebibliography}{16}
\bibitem[{THz()}]{THzPRL}
\bibinfo{note}{Since we are considering incoherent emission we do not here deal
  with the complication that the plasma rod is swept out in time, this being
  reasonable since the emission will not initiate until well after the pulse
  has passed.}
\end{thebibliography}
\end{document}